\documentclass{article}
\usepackage{jim}%,hyperref}
\usepackage[T1]{fontenc}
\usepackage{graphicx}
\usepackage{setspace}
\usepackage{maths}
\usepackage{amsmath}
\usepackage{amssymb}

\usepackage[square, numbers]{natbib}
\title{Creativity in the era of artificial intelligence}

\oneauthor
  {Philippe Esling, Ninon Devis} {IRCAM - CNRS UMR 9912 STMS, Sorbonne Universite \\ \texttt{esling@ircam.fr}}

\begin{document}
\maketitle
\begin{abstract}

Creativity is a deeply debated topic, as this concept is arguably quintessential to our humanity. Across different epochs, it has been infused with an extensive variety of meanings relevant to that era. Along these, the evolution of technology have provided a plurality of novel tools for creative purposes. Recently, the advent of Artificial Intelligence (AI), through \textit{deep learning} approaches, have seen proficient successes across various applications. The use of such technologies for creativity appear in a natural continuity to the artistic trend of this century. However, the aura of a technological artefact labeled as \textit{intelligent} has unleashed passionate and somewhat unhinged debates on its implication for creative endeavors. In this paper, we aim to provide a new perspective on the question of creativity at the era of AI, by blurring the frontier between social and computational sciences. To do so, we rely on reflections from social science studies of creativity to view how current AI would be considered through this lens. As creativity is a highly context-prone concept, we underline the limits and deficiencies of current AI, requiring to move towards \textit{artificial creativity}. We argue that the objective of trying to purely mimic human creative traits towards a self-contained \textit{ex-nihilo} generative machine would be highly counterproductive, putting us at risk of not harnessing the almost unlimited possibilities offered by the sheer computational power of artificial agents. 

\end{abstract}
\section{Introduction} \label{sec:introduction}

Despite its substantial contributions to scientific research, Artificial Intelligence (AI) has focused on \textit{mathematico-logical} approaches, aiming to solve formal problems with a set of well-defined goals. Even with the constant thrilling leaps in this field, recently crystallized around the field of \textit{deep learning}, the current state of our knowledge seems to remain far from understanding \textit{creative endeavors}. The study of this more intricate human behavior proves to be crucial through two main aspects. On the one hand, it involves the understanding of \textit{creativity}, this aspect that so fundamentally distinguishes human beings from other branches of the tree of life \cite{runco2010creativity}. On the other hand, its principal object of study requires to model cognitive phenomena that are central to our social and intellectual evolution \cite{gabora2010evolutionary}. The growing interest for these issues is reflected in the widespread research efforts on \textit{generative models} for a wide variety of tasks from diverse horizons, spanning industrial to fundamental sciences \cite{kobyzev2020normalizing}. %This trend underlines the need to study creative behaviors for future scientific discoveries. 

Amidst these questions, music provides an ideal framework for developing our comprehension of creative behaviors, as it brings together stimulating theoretical questions and cognitive processes that are strenuous to model. Indeed, music operates on unsupervised objectives, hardly defined through task-oriented goals. Hence, revealing these musical creative mechanisms can also provide a remarkable projector on creativity in general, as music is one of the most highly organized, interactive and complex human activity, while being an abstract, sensitive and physical one, profoundly shaping a powerful metaphor of human creative interactions. From an epistemological perspective, approaches in computational creativity can be broadly divided into two major view. Historically, computational models developed for problem-solving aspects have been hijacked \textit{a posteriori} by artists as a form of \textit{creative recycling}. However, within the AI revolution, another line of thought seems to have carved its own path, which we term the \textit{mathematical reification of cognitive hypotheses}. Indeed, AI is strewn by mimicry of intelligent behavior witnessed in biological systems, which appears foundational to this field. We, as humans, are bound to address only ideas that our own perception can reach. As Blake wrote "\textit{Man's desires are limited by his perceptions; none can desire what he has not perceiv'd}" \cite{blake1790marriage}. Hence, we can only reason on conceptual objects that are accessible to our own thoughts. However, this reification approach appears inherently limited, as it is stranded by our current observations and limited cognitive knowledge.

%In this endeavor, we dwell not into asking nor answering scientific questions, but rather on the process of how we should explore these questions as a matter of utmost importance in scientific constructs. 

In this paper, we try to stroll down a novel path of thought, taking a different stance to that of reification, aiming to scrutinize mathematical aspects of AI through the sociological aspects of creativity. Hence, we seek to blur the frontier between social and computational sciences by infusing mathematical properties of AI across a concomitant review of cognitive and psychological studies of creativity. This approach can be seen as an intellectual experiment reflecting our current understanding of creativity in the era of AI. As we try to evaluate the inherent limitations, and what path could be contrived for future researches in AI, we argue that computers have a propensity to expand our limits, and are complementary to our own flaws and limits. Hence, an approach purely based on mimicry of human cognitive behavior would be counterproductive, putting us at risk of not harnessing the almost unlimited possibilities offered by the sheer computational power of machines. Studying these questions could foster and enrich the relationship between humans and AI by targeting situations of partnership converging towards more \textit{symbiotic co-creative} interactions. Hence, addressing these questions could give rise to a novel generic category of creatively intelligent systems.

\section{Epistemology of creativity} \label{sec:epistemology}

The notion of \textit{creativity}, core to this paper reflection, is a topic that could hardly be more central and inseparable from our humanity. In all human crafts, arts and science alike, progress seems to be rhythmed by the pace of our capacity to divorce ourselves with the present, to reinvent and overtake existing thought patterns, and \textit{create} novel ones. While appearing so quintessential and fundamentally distinguishing human beings from other branches of the tree of life, creativity might be the most prominent example of a mental phenomenon that is so central to our own existence, yet we understand so little about it \cite{dietrich2019brain}.

\subsection{Defining creativity} \label{sec:defining}

As we aim to reason on conceptual objects, understanding creativity requires first the ability to precisely define it. However, the concept of creativity appears to be particularly multi-faceted and complex to define. As is usually the case with such highly abstract concepts, decades of debates have first focused on carving out its uniqueness by delineating it from closely-related concepts such as originality, genius, imagination and talent \cite{runco2010creativity}. In the collective subconscious, the core aspect of creativity lies in \textit{novelty}. In that sense, creativity involves moving across our preconceived knowledge and creating a schism between the present and the future \cite{craft2003limits}. Through this first lens, creativity seems to put a strain on our relationship with the future, as it introduces uncertainty. Embracing creativity implies a hazardous leap forward that might upset the balance of our pattern-seeking habits, as it is impossible to fully understand the consequences of an entirely novel concept or object \cite{moran2010roles}. However, we might take comfort in the fact that this uncertainty is not boundless, as creativity (and art alike) does not arise from a conceptual void \cite{locher2010does}. Even though creativity lives in the realms of the least predictable concepts, it remains first and foremost a contextually-embedded phenomenon \cite{lubart2010cross} 

Indeed, creativity involves social aspects, as it implicates individuals in a context, working with a set of existing patterns of meanings and symbols at a given time in history \cite{geertz1973interpretation}. The significance and relevance of creative ideas is not solely observed in their content, but within the social framework and historical period at which these ideas are produced. Ideas appear relevant only when a group of persons articulate their thoughts around the same set of questions \cite{runco2010creativity}, and a critical mass of knowledge develops in one place. %As creativity is a driver of novelty, it appears obvious that novelty itself can only be judged inside a given context. Novelty in a given society might be already moot in another. 
Hence, creativity offers variation and depends equally on the properties of the environment as on its own quality \cite{cropley2010functional}. Conversely, what was deemed revolutionary and obtained widespread acclaim at one point will gradually be integrated as an impending norm of the era, slowly becoming mundane in the society and domain it operates \cite{csikszentmihalyi1988society}. An emblematic example in the musical realm is that of \textit{musical synthesizers} \cite{puckette2007theory}. When they appeared, audio synthesizers remained for long confined to the fringe of contemporary and experimental music, producing unheard sounds that were too unsettling to be deemed musical yet. It is only through a gradual evolution period that these sounds emerged across almost all musical genres, which now commonly integrate different levels of audio synthesis. In that sense, creativity is a transformative process, adapting and integrating elements to a domain with predefined norms, steadily shifting and adapting these norms \cite{puccio2001implicit}.

In summary, although creativity is complex and multi-faceted, it can be articulated around three major components of \textit{novelty} (creative ideas are innovative), \textit{quality} (appeal of the idea) and \textit{relevance} (the idea is appropriate to the task and era) \cite{kaufman2010cambridge}. Hence, studying creativity requires to consider a large number of nuances, which are themselves mostly subjective and renders the analysis of creativity as a set of empirically testable hypotheses rather tedious. This might explain why the empirical study of creativity is only at a very precocious stage. %Hence, most researches have focused on more metaphorical theories, trying to offer speculative explanations to pave the way towards a more broad understanding of its reality (\cite{guilford1950creativity}), or usually targeted rather narrow aspects of its own definition (\cite{kozbelt2010theories}). 

\subsection{Historical aspects}

There is a strong historical component to the development and social perception of creativity, for long stranded between notions of madness or genius. Different eras and societies have been more or less conducive to flourishing creativity, with blossoming periods such as Ancient Greece, Italian Renaissance or French Lumières. To understand these disparities, we provide a scarce outline of the conceptual evolution of creativity along history, based the excellent work of Runco \cite{runco2010creativity}. For a long time, Western societies only considered creative traits in the artistic domain, with a widespread predilection for the myth of the \textit{"lone genius"}. Overlaying the strongly theistic visions of societies at that time, creativity was viewed as a divine intervention, manifested as an outside \textit{"spirit"} or \textit{"muse"} for which the individual creator was merely a conduit \cite{runco2010creativity}. This vision started to shatter, when the scientific era blossomed around the 18\textsuperscript{th} century. This most influential event in the history of creativity, consecrated that all people exhibit different levels of talent in the wake of education, and that the "original genius", which was a form of rule-defying exception \cite{engell1982creative}, was divorced from the supernatural. In response to the industrialization of Europe, Rousseau and the Romanticism movement expressed a separation between the scientific rationalism and the need for humans to rely on their natural feelings as a source of wisdom. This vision created a schism in the societal vision fracturing the rational scientist with the misunderstood deviant artist \cite{runco2010creativity}. This might have created a paradoxical prejudice allowing to denigrate artists and creativity as being confined to deviant personalities. This misconception unfortunately somehow survives in the collective unconscious to this day. Although creative persons and inventors are touted as a driving force to the improvement of society \cite{moran2010roles}, this image of deviance still allows for the denigration of artists when need be.
 
%The birth of social sciences in the 19\textsuperscript{th} century as a mean to understand the "inner world", as opposed to that of physical law propelled questions surrounding the human nature. The seminal contribution to the empirical study of creativity came indirectly from the works of \cite{galton1883inquiries}, questioning the wide diversity that is manifested in different individuals, through broad evolutionary mechanisms. Then, significant approaches to address creativity from a scientific perspective started to flourish (\cite{runco2010creativity}).

%\textbf{9. Further separation between sciences (rational) and arts (deviant)}

%Unfortunately, this created a paradoxical misused argument allowing to denigrate artists and creativity as being confined to deviant personalities. This misconception unfortunately somehow survived in the collective unconscious to this day. Nowadays, creative persons and inventors are touted by leaders as "saviors" for the ills of society (\cite{moran2010roles}), but their image of deviance still allows for denigration when need be.

\subsection{Roles and construction} \label{sec:roles_construction}

%\textbf{1. What is the role played by creativity in society} - \textbf{2. Why does it matter}

A major question lies in the role that creativity plays in a given society, and \textit{why} creativity might be so important. This outlines the perceived impact of creativity, but also what aspects are valued in creativity. %, as evaluated by experts of the field in a given context. These evaluations usually are biased towards producing works that are similar to the existing, and radically novel ideas usually need a longer term development before being accepted as a novel norm. 
As discussed by Gardner \cite{gardner1993creating}, creativity can be seen as a temporary misalignment between an idea and the society in which it develops. Eventually, as some people are willing to take more risk and embrace new ideas, these gradually integrate the social fabric and ultimately become an accepted and standardized part of society. 

\subsubsection{Creativity in society} \label{sec:society}

%\textbf{1. Moran definition of the two roles of creativity, opening to a third one.}

Moran \cite{moran2010roles} proposed to study the functional aspects of the \textit{role} that creativity could play in society, proposing that it mostly endorses functions of \textit{improvement} or \textit{expression}. The \textit{improvement role} is the effect that a creative object can have on society (as technological artefacts), while the \textit{expression role} is focused on the role that creativity exert on an individual. Although these two roles seem somewhat dichotomous, they are proposed to interact in more of a complementary than competitive way. \\

\textit{2.3.1.1. Creativity role as improvement}

The vision of creativity as \textit{improvement} is that which is often glorified by political or industrial leaders, as a mandatory tool for the advance of society and humanity. In that sense, society is seen as a system, which is constantly moving upwards to an hypothetical blessed state (notwithstanding our complete absence of a shred of knowledge to where that might be). Hence, creativity is the tool that transport society across the borders of the present towards an idealized future, by shifting norms to a higher position \cite{runco2010creativity}. Whereas practitioners develop norms, creativity pulls society forward, while inevitably giving way to standardization \cite{moran2010roles}. In that view of creativity as an enhancer of society, its importance is the goal and that it allows us to progress towards it. \\

\textit{2.3.1.2. Creativity role as expression}

The other role of creativity proposed by Moran \cite{moran2010roles} is that of \textit{expression}, which might be interpreted differently depending on the society in which it unfolds. In that aspect, creativity can be regarded as a mean of self-expression and individuality, while the exact nature of this expression depends on the permissiveness of the surrounding society. These forms of creativity in society are less focused on their results, and rather allow for solipsistic and individualistic approaches. Hence, creativity as expression can be seen as a cathartic activity, which might be less valued by society.\\ \\

\textit{2.3.1.3. Creativity as transformation of expertise}

The two roles of creativity in society can interact and produce, for instance, improvement as a result of self-expression. In any case, it is important to note that creativity is before all a \textit{transformational} activity. There is a widely agreed-upon consensus that creativity is only permitted through an existing body of knowledge, strongly influencing the natue and quality of creative outcomes. As stated previously, the uncertainty of creativity is not boundless, as nothing arises from a conceptual void. When Newton stated "\textit{If I have seen further it is by standing on the shoulders of giants.}", he recognized that his own discoveries were only made possible by all the knowledge accumulated by previous researchers before him. Hence, creativity is the result of community-built expertise, later \textit{transformed} in an incremental fashion%, expressed in the \textit{evolutionary} theory of creativity. % This is indubitably the result of both individual and social factors \cite{seitz2003communitarian}, which acts together.

\subsubsection{Evolution and construction} \label{sec:evolutionary}

This transformational view is epitomized in the \textit{evolutionary} theory of creativity, which proposes an interesting parallel between genetic evolution and the development of creative ideas. The original Darwinian model of Simonton  \cite{simonton1999origins} aimed to describe more developmental aspect of the creative process, but it naturally extends to larger (societal) scales of how creativity developed across time and how social factors can come into play. Indeed, creative ideas are built on previous ones in an adaptive and open-ended manner \cite{gabora2010evolutionary}. This view of creative ideas evolving over time through culture can be depicted in Darwinian terms \cite{aunger2000darwinizing}, as pertaining to a form of \textit{inheritance of ideas}, which are incrementally adapted to the timely constraints of their social environment as they pass from one person to the next. This concept is termed as the \textit{dual-inheritance theory} \cite{gabora2010evolutionary}, which emphasizes the fact that we inherit both biological but also cultural information. This theory views culture as discrete elements, which are submitted to an adaptation process, both composed of random mutations and a fortuitous process where ideas are selected because of outside environmental effects \cite{dennett1988conditions}. 

In that sense, we explore an unknown space of ideas (\textit{variation}) and choose to pursue some and not others (\textit{selection}), turning creativity in a variation-selection algorithm informed by expertise \cite{dietrich2019brain}. Here, the social context is of prime interest, as creative ideas are observed equally for their content and within the social framework within which these are produced. As ideas appear relevant only when a critical mass of knowledge is articulated around similar thoughts, this further underlines the ubiquitous need for knowledge, which is transformed through creative processes. However, this thickens the complexity of clearly separating intelligence and creativity. Indeed, intelligence can arguably be described as the process of associating and transforming existing knowledge. Yet, as social scientists debated for over a century to delineate these two behaviors, this warrants the existence of a similar need for moving our reasoning from artificial intelligence towards artificial creativity.

\subsection{Towards the existence of artificial creativity} \label{sec:existence_of}

Being able to truly delineate intelligence and creativity would mean that they conceptually live as independent \textit{orthogonal} dimensions, which can each be evaluated separately from the other. Hence, this poses the question of the \textit{measure} and \textit{evaluation} of creativity as a dimension dissociated from intelligence.

\subsubsection{Evaluating creativity} \label{sec:evaluating}

The question of assessing creativity might be one of the most controversial and complicated issue of the field. This approach (called \textit{psychometric}) is unique as it also radiates and impact all other types of creativity studies \cite{kozbelt2010theories}. This entails the sensitive question of \textit{measurement}, which appears at first to be dauntingly complex in creative fields. Yet, this field has a very extensive and flourishing research history \cite{plucker1998death}, entailing the \textit{reliability} (agreement and consistency), \textit{validity} (accuracy of the measurements) \cite{kozbelt2010theories}, and \textit{discriminant validity} (not being contaminated by correlated concepts of intelligence) \cite{wallach1965modes}.

Most approaches to evaluating creativity have revolved around the idea of separating between \textit{convergent} and \textit{divergent} thinking processes. The \textit{convergent thinking} process (a term coined by Joy Paul Guilford \cite{guilford1950creativity}) corresponds to the use of knowledge and reasoning to solve a problem by eliciting a \textit{single solution}. Hence, this delineates a corresponding set of questions that have a single \textit{correct} answer. Oppositely, the process of \textit{divergent thinking} requires a framework where a wide variety of ideas can be generated in response to a given question or stimulus. These tests are usually derived from the seminal Torrance Test of Creative Thinking, which allows for the construction of associative hierarchies, such as asking to propose different novel uses of a commonly known object \cite{dietrich2019brain}. In the musical realm, another stream of research rely on improvisation in jazz where a given melody should either be completed from memory (control) or freely (creative), while keeping some parameters (length, tempo) fixed \cite{plucker2010assessment}.

Interestingly, this dichotomy would appear to provide a clear-cut and simple solution to our previous conundrum. By crudely exaggerating, we could conclude that intelligence is convergent, while creativity is divergent. However, it should also be noted that some criticisms exist on the fact that creative processes also result from convergent thinking \cite{dietrich2019brain}, as creative ideas can be the fruit of laborious trial-and-error works. Still, the domain of \textit{divergent thinking} seems to remain a privileged ground reserved to creative behaviors.

\subsubsection{AI for creativity in the light of social sciences} \label{sec:ai_in_social}

Following our previous discussions, it would first seem that AI approaches are only relevant to convergent thinking. Indeed, it is usually secluded to the definition of \textit{problem-solving} approaches, where the learning is formalized by having a given \textit{single correct solution}, which allows to understand and evaluate the quality of the proposed model. However, these limitations are partly bypassed in the AI domain by using approaches based on probabilistic formulations \cite{bishop2006pattern}, where we try to understand the entire \textit{distribution} of given types of data, rather than answering a question (such as classifying objects).

If we take a closer look at the \textit{criteria} that are used for divergent thinking tests, these are usually evaluated on several indicators measuring the \textit{ideational fluency} (amount of answers), \textit{originality} (unusualness of answers), \textit{flexibility} (variance in the concepts elaborated) and \textit{elaboration} (precision and details of the answers) \cite{plucker2010assessment}. A criticism of this evaluation method is that fluency can act as a contaminating factor in the originality scores, emphasizing the quantity over quality of ideas \cite{plucker2010assessment}.

By construction, probabilistic modeling could easily overpower any human on these tests for \textit{fluency} and \textit{flexibility}. Indeed, an approach based on probability distribution estimation can produce an infinite number of solutions and these can be as distant as we want. However, there is two caveats to this reasoning. First, this implies that we rely on a random sampling process to draw solutions, which already means that we perform some choice on the generative process. However, this generation being an (hypothetically) entirely random decision mechanism, it can already be seen as a form of human choice on the creative process. Second, the flexibility aspect is related to the sampling of a \textit{pre-defined} distribution, which limits the model to produce examples that remain consistent with its original observations. Therefore, this greatly limits the output of the model to the set of knowledge that was selected and provided at the onset.

As discussed earlier, expertise is a fundamental part of creativity, as it allows to obtain more efficient reasoning, based on appropriate problem representations, and recalling domain-relevant patterns or invariant characteristics \cite{kozbelt2010theories}. Following our ideas from the previous sections, it also seems that this social context component in the evaluation of creativity can be limiting to human creativity. Oppositely, computational approaches can be freed from these societal biases, and generate an almost infinite number of solutions. However, this also introduces a novel caveat, that AI is trained to optimize a given distribution of pre-existing knowledge and it cannot evaluate the quality of its own solution in any other way than the criterion, which is provided for training it. This lead to a paradoxical situation, where there can be a boundless number of solutions, but in an inherently limited system bound by the model, knowledge and criterion choice.

All of these observations warrants the question of what could be the place of AI in the creative process. Hence, we loop back to our previous question of finding a way to transition from artificial intelligence to artificial creativity. Indeed, if we have access to a seemingly infinite generation machine, but this machine is unable to evaluate its own correctness, and only produces variations of existing concepts, what creative use could we make of this system with peculiar characteristics.

\section{On the intrinsic limits of AI for self-contained creativity} \label{sec:limits}

We now try to cast a light on the limits of relying on AI to produce its own self-contained creative behaviors, by discussing the main body of our argument, through reviewing all studies surrounding the creative \textit{process}.

\subsection{Studying the creative process}

The dominant paradigm to study the creative process scrutinizes its structure as a set of \textit{stages}, defining \textit{componential} cognitive processes in a sequential or recursive fashion. The seminal model of Wallas \cite{wallas1926art} separates the creative process between the \textit{preparation} (information gathering), \textit{definition} (problem finding), \textit{incubation} (reflecting on ideas), \textit{illumination} (appearance of a solution) and \textit{verification} (testing the quality of the solution) stages. This very linear model has since been more widely replaced with a cyclical one, where stages are performed in various combinations and are highly influenced by (extrinsic or intrinsic) motivation \cite{runco1995cognition}. Although it is now recognized that creativity is dynamic and interactive, looping between different stages, the original components of Wallas remain widely used even in modern refinements.

In the view of cognitive theory, any phenomenon can potentially be recreated if we understand all of its principles. This should allow to address any complex and arduous problem by splitting it in smaller understandable phenomena. The underlying hypothesis is that ill-defined problems can be broken into a set of smaller well-defined objectives \cite{simon1996sciences}. This cognitive approach, which focuses on the operators and strategies that come into play in the creative processes has been a privileged playground for defining how AI should mimic creative behaviors.

\subsection{Mathematical reification of ideas}

The abstraction of the creative process has been a core study for cognitive approaches and one of the most fertile ground for the \textit{mathematical reification of ideas} in AI research. In that view, AI should be bound to mimic human creative endeavors by reproducing the abstraction and tactics that come into play in our brains. In that sense, the cognitive approaches have been mostly concerned with fundamental processes that can be translated in computational terms. This entails the extent to which \textit{knowledge and information} are organized and accessed, through different types of memory systems, and the corresponding processes for \textit{retrieval and analysis} of various sources of information. Subsequently, the operations applied to this knowledge allow to circumscribe the creative process in a computational sense. Here, we broadly separate these views between these \textit{information} and \textit{operation} aspects, with some permeability across these categories. We study the major sub-processes as they have been analyzed in cognitive studies, while trying to tie links with how this reflects on corresponding approaches in AI research.

\subsubsection{Information}

There is a widely agreed consensus that creativity is only permitted through an existing body of knowledge, which strongly influences the quality of creative outcomes. This influence can be analyzed through the ways this knowledge is \textit{organized} (information structures), \textit{retrieved} (access strategies), \textit{analyzed} (similarity evaluation) and eventually \textit{transformed}. \\

\textit{3.2.1.1. Organization} \label{sec:organization}

When receiving novel knowledge, humans create information structures and hierarchies allowing to memorize and organize it. At the neural level, it appears that neighboring neurons encode similar micro-features, and distance between neurons can be interpreted as a proxy for their feature similarity \cite{gabora2010evolutionary}. Furthermore, these distributed representations of micro-features induce a natural modularity that do not require any proactive mechanism for high-level organization. As the memory supposedly work in a content-addressable way, it appears reasonable that we gradually transitioned from coarse to finer representations of memories \cite{gabora2010evolutionary}. Hence, we evolved towards increasingly complex overlapping distributions, allowing for more interconnections and recall, leading to relationships drawn from more integrative internal representation.

These more complex associative hierarchies allow to handle intricate concepts with a variable organization of cognitive elements, objects and relationships \cite{kaufman2010neurobiological}. The appearance of these systems is theorized to have produced emerging complex internal representation of abstract meanings through symbols and their relationships. As proposed by Deacon \cite{deacon1998symbolic}, we shifted from an \textit{iconic} representation (simply storing physical or visual properties) to an \textit{indexical} representation (representing a set of properties) and finally to a \textit{symbolic} representation (where the representation itself bears no similarity to the object it represents). For Deacon, the birth of symbols allowed us to imagine the use of objects as elements with correlations separate from their simple physical properties, enabling more intuitive and associative thinking processes. All of these ideas are crystallized in AI around the \textit{representation learning} field \cite{bengio2013representation}, where the goal is to understand and learn properties of a given set of objects in an \textit{unsupervised} way (only having access to the objects themselves). This is usually performed based on some more or less complex forms of \textit{compression} in order to produce higher-level and smaller representation hierarchies. \\

\textit{3.2.1.2. Retrieval}

Parallel to the development of these specialized neural circuits to structure knowledge, our brains also required to define mechanisms to access and retrieve this information. Indeed, when engaging in creative thinking, we need to rely on different granularities of representation levels, from very specific to highly abstract. The resulting retrieval process obviously depends on the underlying organization but also the \textit{accessibility} of different knowledge items \cite{ward2010cognition}. The internal (physical) constraints of retrieval can provide either highly similar ideas, or oppositely aim to process abstract and distant concepts.

An encompassing model for understanding cognitive information retrieval is the \textit{path-of-least-resistance} proposed by Ward \cite{ward1994structured}. This model posits that our predominant retrieval mechanism is to access basic and specific low-level examples from a given domain as starting points, while projecting their properties on the novel task at hand. This suggests that we can use the representativeness of information items as a retrieval likelihood function. It also supports the idea that remote associations, less representative and distant concepts can lead to higher novelty \cite{ward2010cognition}. Supplementary constraints such as \textit{latent inhibition} allow us to attend selectively to those information that appear the most relevant, while screening out irrelevant knowledge. However, it has been shown that the lack of such inhibitions might provide a greater ideational fluency of creativity \cite{feist2010function}. Indeed, using more distant concepts have been shown to increase \textit{originality}, but this may come at the cost of \textit{practicality}. This process of contextually selecting some parts in information is also highly reflected in modern AI through the idea of \textit{attention} mechanisms \cite{vaswani2017attention}. These approaches compute additional vectors that allow to contextually mask part of the information at different levels of processing, leading to the now well-known \textit{transformers} models. \\

\textit{3.2.1.3. Analysis}

Our ability for organized information retrieval allows to perform further \textit{analyses} of this knowledge. This process has been scrutinized through the notion of \textit{contextual focus}, where creative inspiration occurs when our attention is defocused, allowing more associative thoughts and to activate simultaneously distant representations \cite{martindale1999biological}. This form of \textit{defocused attention} would allow to broadcast diffusely to broader region of our memory, allowing to consider a larger variety of elements rather than attentively select distinct elements as in convergent thinking. This type of broad activation with looser definition of similarity and selection could be primordial to associative thinking, while still being based on probabilistic relationships between knowledge elements. This posits the existence of a separation between \textit{explicit} cognition maps allowing convergent reasoning for problem-specific approaches, while \textit{implicit} cognition would allow to reason on distant creative associations between elements \cite{reber1989implicit}. Interestingly, these hypotheses contrast with our previous parallel to the \textit{attention} mechanisms in AI, pointing out to a potential paradoxical limit in this reification process. \\ %This is also related to the notion of \textit{processing fluency}, which states that our aesthetic response to stimuli depends on how easy it is for us to perceive the input on high level properties such as proportion, symmetry and complexity (\cite{aleem2019beauty}). \\

\textit{3.2.1.4. Transformation and meta-representation}

Although transforming information could already seem to belong to the \textit{operations} category, reorganization and transformation might already occur on our existing information structures. Indeed, as the phenomenon of neural plasticity is commonly accepted, there is also potentially some reorganizing mechanisms to allow existing knowledge categories to be formed based on task-specific objectives \cite{ward2010cognition}. This could be facilitated by modules of \textit{meta-representation}, serving to represent "concepts over concepts" \cite{sperber1994modularity}. This interesting notion would permit to facilitate high-level reasoning across different domains, especially in the case of \textit{analogies} and \textit{metaphors}. Hence, this type of second-order system could provide us with the ability to reflect on our own knowledge representations and processes \cite{dennett1988conditions}. \\

\subsubsection{Operations}

Cognitive approaches emphasize the concept that individuals generate creative ideas by exploring their knowledge through different types of operations. Hence, we discuss in this section the ways in which we explore and develop relationships between different knowledge items. Several attempts to produce an encompassing model of such creative operations have been proposed, such as the \textit{propulsion} model of Sternberg \cite{sternberg2002creativity}. This model delineates a classification of creative operations as \textit{replication} (transforming known ideas), \textit{redefinition} (seeing known ideas in a new way), \textit{incrementation} (extending known ideas), \textit{advance incrementation} (similar but going further an acceptable threshold), \textit{reconstruction} (reviving a previously abandoned idea) and \textit{reinitiation} (starting an idea at a radically different new point). Although this proposal is interesting, we rather follow here a categorization which is closer to the technical AI-based views and more prone to potential reification. Hence, we split different ideas between the \textit{modification} (transforming a single item of knowledge), \textit{association} (linking several items or ideas), \textit{analogy} (existing ideas are transferred to a new domain) and \textit{abstraction} (finding a more general concept encompassing several existing ideas). \\

\textit{3.2.2.1. Modification}

The large majority of advances in any domain are based on small incremental changes in various ideas pertaining to that field. These operations imply to slowly \textit{modify} existing ideas to include increasingly complex variations. This appears logical as creative objects must usually strike the right balance between familiarity and novelty. This behavior can also clearly be seen in the publication patterns of any scientific field. This operation can also sometimes be evidenced even within the production of a single object. For instance, in music, the notion of \textit{theme} and \textit{variations} is highly present, where the same material is exposed and varied along a given musical piece. With this modification operation, it is usually necessary to understand the existing similarity with the target idea and features inside the given domain. This would imply a gradual access to the most readily similar instances given similar features, leading to a form of exemplar generation \cite{ward1995s}. Although this can be seen in several approaches in AI, the variations remain in a confined set as delineated by the choice of the model and dataset, following our previous argument pointing out to this inherent limitation (Section~\ref{sec:ai_in_social}). \\

\textit{3.2.2.2. Association}

The second type of well-studied creative operation is that of \textit{association} or \textit{combination} of existing knowledge and ideas. The conceptual combination implies to process complex forms of similarities between apparently heterogeneous ideas, in order to generate a novel concept by merging these elementary ideas. This notion of allowing two contradictory ideas to be entertained simultaneously, called \textit{Janusian thinking} is critical to creative ideation. Usually, this combination process allows the emergence of features that are beyond the simple summation of their components. Hence, combining distant concept and reconciling their discrepancies allows to postulate novel properties that are absent from the original concepts, which is core to creativity \cite{ward2010cognition}. Generally, the combination process requires to perform integration between ideas that are not usually grouped into a single coherent concept or even space, which requires to break traditional thinking patterns. Regarding AI approaches, this seems extremely complex, as it transgresses the notion of physical properties distance, and even requires to go beyond conceptual distance aspects. To the best of our knowledge, this operation appears to currently be out-of-reach for AI models. \\

\textit{3.2.2.3. Analogy}

Another type of process that has been repeatedly examined in creativity studies is the concept of \textit{analogy}, where existing well-structured knowledge is projected onto a novel domain. On a simple basis, analogies can be used to apply given solutions from a domain to another, or communicating ideas in a more concise way. However, the real force of analogical reasoning appears when it connects the source and target domains at very profound levels of knowledge, rather than merely on their surface \cite{gentner2001analogical}. As an interesting but somewhat looser parallel, this notion is currently highly studied and successful in the AI field, defined as the task of \textit{domain transfer} \cite{bitton2018modulated}. In this approach, part of the properties from an object (defined as \textit{source content}) are transferred to a given domain (defined as \textit{target style}). However, this process mostly aims to transform some low-level perceptual properties of objects rather than some profound conceptual aspects. \\

\textit{3.2.2.4. Abstraction}

The \textit{abstraction} category is the most intricate and complex to grasp, as it involves to work at higher-level spaces of logical reasoning. Here, we link this idea to our inherent \textit{predictive} system that is thought to be one of the critical function of the brain \cite{dietrich2019brain}. Evidence suggest that there are general appraisal and reward-based mechanisms in the brain that would provide incentives for learning. In that sense, the abstraction operation could allow us to perform more accurate predictions, by using these past rewards as global indicators for enhancing our own decision mechanisms. This relates to our previous discussion on the definition and learning of novel representation spaces (Section~\ref{sec:organization}). Indeed, some approaches have tried to perform prediction directly in these spaces \cite{oord2018representation}, as a proxy to reorganizing knowledge. However, these still do not provide forms of higher abstraction.

\subsubsection{Limitations}

As we reflect back on our presentation of different aspects of the creative process, we can see that our understanding of different stages is limited to that of \textit{preparation} (seen as information gathering) and \textit{incubation} (seen as operations on this knowledge). Following various components, there seems to be currently no equivalent to the steps of \textit{problem finding} and \textit{illumination}. However, these steps are critical to the creative process, and it appears that the most functional parts of any purely creative behaviors are still entirely human-based. This would point out to a form of partiality in artificial creativity, as the major aspects of its stages still seem to be out of reach.

As we outlined throughout the definition of creativity (Section~\ref{sec:defining}), the appreciation of its characteristics is usually separated between the three factors of \textit{novelty}, \textit{quality} and \textit{social relevance}. Hence, we try in the following to delineate the potential limitations of AI for creativity based on these three major criteria separately.

\subsection{Creativity and novelty}

The \textit{novelty} aspect of creativity might be one of the most prominent in the collective subconscious. However, this might also be one of the most critical and complicated point to address in a computational approach. This relates to the way that systems are optimized by learning on a set of examples \cite{goodfellow2016deep}. Indeed, training a given model usually relies on minimizing the \textit{expectation} of a loss (error) term. This implies that we are computing a \textit{mean} error across a set of known examples. This also means that the model is incentivized to perform correctly on the \textit{most common} elements of knowledge, by trying to fit the \textit{principal mode} of the distribution (where most examples are concentrated). Conversely, the unusual (outliers) examples will have the lowest impact on the model training, and will mostly be ignored to avoid skewing the \textit{global error term}. However, as we discussed earlier, creativity highly relates to dealing with these types of distant examples. Hence, the major question is to know whether AI models could truly provoke novelty through some given operational mechanisms, or if this would be bound to require a human intervention. Although AI is successful in the organization and retrieval of knowledge, the combination of conceptual elements drawn from memory stem from largely more complex processes.

The psychologist Margaret Boden has given much attention to the relations between creativity and machines \cite{boden1998creativity}. In this view, the ability to find new, surprising and socially valuable ideas can occur in three main ways: \textit{combinatorial} (producing novel configurations of familiar materials), \textit{exploratory} (discovering new paths in conceptual spaces) or \textit{transformative} (when the space itself is disrupted giving way to ideas that were previously inconceivable). Hence, there might be some yet uncovered operations that might favor creativity in AI, by empowering their \textit{decision-making} traits. This line of thought is principally developed in the \textit{reinforcement learning} approaches, where agents are defined to explore spaces of possibilities. However, these require the definition of rewards and success functions that are still complex to define, as they relate to subtle perceptual and contextual aspects of the generated objects.

\subsection{Evaluating the creative quality}

Regarding the evaluation of the creative \textit{quality}, there appears to be a strong duality in this question. On one hand, learning approaches are trained by minimizing a criterion, which should serve as a proxy for quality. However, this loss usually pertains to some \textit{structural} aspects of the generated data, rather to more abstract aspects of its content. Hence, AI models are still unable to determine the real \textit{creative value} of what they produce. To allow this complexity of understanding, we need to be able to evaluate both the creative \textit{product}, but also the \textit{process} itself.

Hence, one of the major flaws of AI applied to creativity appears to lie within its inability to judge the creative structures that emerge from exploratory processes. In a more global sense, this amounts to say that AI approaches bear no \textit{artistic intent}. Although this appears to be a pretty straightforward observation, this can be traced back to the question of \textit{measurement}. Indeed, in order to provide a solution to this issue, we would need not only a computational definition of creative ideas, but most importantly a \textit{criterion} of evaluation on what are low-probability events that still appear relevant. This question of relevance relates strongly on the notion that creativity can only be evaluated within a given societal context.

\subsection{The notion of context}

As we emphasized in previous discussions, creativity is highly contextual and deeply related to cultural aspects. One of the most notable differences in the conception of creativity can be found in the separation between Western societies seeing creativity as a godly intervention conveyed through an outside spirit, and Eastern societies that historically considered creation as a discovery or mimicry of something already pre-existing in the universe.  This discrepancy in the evaluation of creativity can also be seen in what different cultures revere as a form of artistic expression. For instance, Japanese civilization highly values the works of calligraphers, sword makers and ceramicists as art \cite{simonton1997foreign}, which is less prominent in Western societies. This influence of culture on our daily life is so pervasive and deeply intertwined with our reasoning that we are not even able to measure the true impact it may have. An emblematic example can be experienced in the excellent book \textit{Flatland} by Edwin Abbott Abbott, showing that a person living in a two-dimensional world will most probably never be able to reason about cubes.

Throughout these aspects, cultures are inherently linked to languages, as they represent both the way we communicate, but also the way we think about concepts. Hence, individuals with diverse backgrounds process information differently, and have vast varieties in conceptual functioning \cite{ward2010cognition}. These differences in language can be generalized to their largest separation between \textit{taxonomic} or \textit{thematic} ways of conceptualizing ideas \cite{markman1984children}. The \textit{taxonomic} functions of languages is highly dominant in Western cultures, where the objects and reasoning are "decontextualized". Hence, the relationships \textit{between} objects are less important than the category to which the object belongs to. Conversely, the \textit{thematic} language constructions which define Eastern languages put a larger emphasis on the context and logical or causal relationships between objects, rather than the individual object taken in isolation. Interestingly, this is also highly reminiscent and symmetrical to the theological ramifications between Western and Eastern constructs, which seem largely impacted by our \textit{logos}.

Hence, it might appear that the evaluation and novelty of creativity remain entangled contextual-prone questions. In order to provide an alternate path to the use of AI in creativity, this requires to redefine the relationships between human and machine in creative endeavors.

\section{Redefining the relationship with AI through co-creativity}

As we have seen across the previous sections, there are currently still some profound limitations to attain a true form of artificial creativity. Indeed, current systems do not show any substantial musical creativity, lacking machine musicianship (the capacity to process music at a structural, symbolic level, a term coined by Robert Rowe \cite{rowe2004machine}), high-level interaction strategies or generative autonomy. However, there still exists interesting avenues to truly harness the peculiarities of AI models in creativity.

\subsection{Different views of AI in creativity}

One crucial distinction needs to be made in the use of AI for creativity depending on the role and place that it is granted in the creative process. On the one hand, a whole body of research appears to be devoted to use AI as \textit{self-sufficient generators}, which seek to generate entire pieces of art. The system receives little input from the user, which now acts as a passive admirer of the creation. On the other hand, AI systems can be seen as \textit{creation tools}, which allows to enhance our own creativity. In that regard, despite its higher complexity and intellectual ramifications, AI simply holds a position akin to that of an evolved brush to a painter. This second view provides enhanced and facilitated ways to explore information spaces, but remain entirely subservient to our own creativity.

As we discussed earlier, when using AI in generative processes, we still need to have a precise and well-defined \textit{problem}, along with a set of representative data. Hence, this already highly constrains the potential behavior of the algorithm, as we delineate the world of possibilities that are attainable and explored by the model. Furthermore, we also implicitly limit the capacity of the models to evaluate their own realization, as we define a certain training criterion and loss function, which acts as a learning signal. However, this will also constrain the view of the model to a single facet of the produced artefact. All of these observations relates to the question of creativity as a \textit{problem-finding} rather than \textit{problem-solving} activity. It is sometimes more crucial to find an interesting question rather to find solutions to existing ones.

Current models of computational creativity focus on the ability to generate novel content. However, as discussed previously, the highly context-prone aspects of creativity seem to confine this \textit{singularity-seeking} approach to a pointless endeavor. Oppositely, it should be recognized that creativity traits should be shared between human and machines. Indeed, most works do not consider the rich interplay and collaborations that emerge in the \textit{interaction} and \textit{control} of creative approaches themselves. Indeed, most models are \textit{proposal generators}, where we can pick from a list of ideas. Instead, we should aim to establish \textit{co-creativity partnerships} focusing on the interactions dynamics, by evaluating a variety of collaboration strategies. To do so, the first step is to define the problem space itself, which can already be a daunting task.

\subsection{AI as a creativity-enhancing tool}

Whatever type of AI we might define, it still requires some form of human supervision. First, relying on AI as a creativity-enhancing tool requires to augment computational creativity models with artificial perception, by learning representation spaces for on-line music structure discovery and generative decisions based on cognitive dynamics by infusing social sciences in computing models. Drawing an analogy to the previous section, first finding \textit{disentangled} representation spaces of information is a decisive and primordial aspect to any creative process.

This process of unsupervised learning is still poorly understood, even though it is one of the key towards new generations of learning algorithms. These questions are crystallized around the field of \textit{disentangled representation learning}. The idea of \textit{representation learning} is to find compressed spaces (termed \textit{latent spaces}) of information with a logical organization, while the \textit{disentanglement} seeks to have the dimensions of these spaces to represent each factor of variation in the data separately. This question is usually approached through the use of Variational Auto-Encoders (VAEs) \cite{kingma2013auto}. Although the use of VAEs for audio applications has only been scarcely investigated, we recently proposed a perceptually-regularized VAE \cite{esling2018generative} that learns a space of audio signals aligned with perceptual ratings via a regularization loss. The resulting space exhibits an organization that is well aligned with perception. Hence, this model appears as a valid candidate to learn an organized audio space. Following this research, we performed several follow-up studies to assess different generative aspects of these models, such as adding musical conditioning \cite{bitton2019assisted}, introducing other modalities of knowledge \cite{chemla2019cross} or providing a continuous drum sounds synthesizer \cite{aouameur2019neural}. These types of spaces can provide an invaluable step in the organization of information for the creative process, and can already serve as a generative tool to quickly explore high-level properties. An exciting direction of research would be to model more accurately the geometric and topological features of these latent spaces across different learning settings. This approach could provide insight on the learning process and new mechanisms to perform unsupervised learning.

These recent researches on information organization link to our question of expanding the human knowledge. Indeed, the strength of AI lies in its ability to understand multivariate and highly non-linear interactions in spaces with large number of dimensions. Hence, this aspect of \textit{knowledge structuring} can enhance our own power over information structures. As a clear pragmatic example of this idea, we recently introduced a radically novel formulation of audio synthesizer control \cite{esling2019universal}. We formalized this as the general question of finding an invertible mapping between organized latent spaces, linking the audio space of a synthesizer's capabilities to the space of its parameters. This model allows to address extremely complex tasks such as \textit{parameter inference} (finding the parameters of a synthesizer that reproduces any audio file), \textit{macro-control learning} (disentangling and simplifying its complex controls), and \textit{audio-based preset exploration} (navigating sets of examples in a perceptually-relevant and intuitive manner) within a single model. This particular instance shows that AI can be used as a mean to efficiently re-organize intricate knowledge, even though it is already accessible in a more entangled and harder-to-harness sense. Hence, AI can be an extremely powerful \textit{creativity facilitator} in its ability to organize complex knowledge in a simplified way.

In order to decisively move towards artificial creativity, we critically need AI models that are able to listen and apprehend the musical structure correctly. These higher-level cognitive phenomena involve long-term memorization and structure discovery, usually performed on multiple time scales. In this direction, we worked on different aspects allowing to perform online semantic information extraction \cite{carsault2018using} and multi-scale musical structure prediction \cite{carsault2019multi}. However, this also warrants the need to perform a \textit{machine evaluation} of human creative behaviors in order to understand and match the complexity dynamics of musical interaction. This would allow artificial entities to infer new modes of a distribution in a self-supervised way as a proxy for creativity. In that sense, an interdisciplinary approach is mandatory, where social science and anthropological studies need to collide with machine capacities and algorithmic approaches.

\subsection{Towards models of co-creativity}

Overall, it seems that the limits of human creativity are reciprocally limiting to AI, as we are bound to define processes that only mimic our own preconceptions of creativity. Furthermore, there is a limit in the quest of self-contained generative AI, as we require these approaches to generate novel content, but yet strongly conform to the existing norms of that domain. A solution to this conundrum warrants the need to profoundly redefine the creative relationship that may exist between human and machines. Indeed, instead of focusing on the either very subservient view of AI as proposal tools, or oppositely seeing it as a self-contained generator, we should acknowledge that the true power to be harnessed comes from the \textit{partnership} between two separate systems (human or machine) with each its specific characteristics. This idea of \textit{co-creativity} emphasizes the fact that creativity is an emerging phenomenon resulting from complex interactions and feed-backs between actors involved in a creative process. This allows to regard not only the production of each (human or artificial) part in isolation, but rather focus on the \textit{interaction} that emerges between different components.

Recently at the IRCAM STMS lab, Assayag \cite{assayag2016improvising} has thoroughly investigated artificial creativity issues involved in musician-machine interaction, especially in the case of co-improvisation where agents of different nature (artificial or human) perform together in highly unpredictable live settings \cite{assayag2010interaction}. Assayag proposed the concept of human machine co-creativity in music \cite{assayag2020human}, in order to overcome the aporia inherent to the essentialization of machines when we try to assign them anthropocentric features such as creativity or intelligence, focusing then more on the relations than on the intrinsic qualities of agents. Co-creativity in that respect addresses emergent distributed behaviors inherent  to complex cross-learning feed-backs between agents, and suggests reinforcement mechanisms pertaining to co-action. This shift in objective allows us to question different ways to build the best possible technical tools that could foster and enhance co-creative interactions. According to Assayag \cite{assayag2020reach}, co-creativity can only appear when two features linked to emergence and non-linear dynamics are identified: (1) emergence of cohesive behaviors that are not reducible to, nor explainable by the mere individual processes of agents; (2) apparition of non-linear regimes of structure formation, leading to rich co-evolution of creative forms. The underlying assumption is that such surging phenomena result from cross-learning processes between agents involving complex feed-backs loops and reinforcement. As a major consequence, the states and behavior of participating agents are in return modified continuously, making them evolve in terms of knowledge and skills. These ideas involve intriguing prospect of modeling the dynamics of different actors that cooperate to perform complex adaptation. This requires anticipatory systems of interaction, with real-time adaptation that could account for the collective dynamics in creativity.

\subsection{Through the doors of perception}

As we discussed earlier, AI is inherently limited to optimize a \textit{mean accuracy} on a given set of data, leading to the perilous situation of being an infinite norm-generating machine, which would be detrimental to creativity. This underlines the risk of overusing AI models towards the dangerous path of standardisation. While we will not dwell into the questions of economic gain that could be intertwined with this prospect, we still underline the dangers of indulging in these goals. The current supremacy of having highly conform thoughts and products, even in the artistic domain, can only lead to a dampening of creative endeavors. As an unconventional analogy from biology, this would resemble a form of \textit{genetic drift} in creativity which should be avoided at all costs. This phenomenon is caused by continuously selecting a given subset of individuals, which causes the reduction of genetic variance and an overall impoverishment of the available genetic material. To transpose this in optimization terms, we would constrain ourselves to converge and remain stuck in a local minimum, by removing large parts of our search space.

Although these thoughts might appear as a bleak observation on the use of AI for creativity, there are glimmers of hope hiding underneath. The advent of computational approaches have propelled creative endeavors into horizons that could never have been reached before, and this path seems far from being exhausted. These exciting avenues come from the fact that AI models and human behaviors can compensate for their respective limitations. In that sense, AI can easily process complex and multivariate information without any prior bias. Hence, it can allow us to deal with objects and concepts that would easily exceed our perception potential. This also question the epistemological reasoning of our quest to perform biologically-inspired mimicry, when we could harness the power of a truly novel and somewhat boundless generation tool.

\section{Conclusion}

Across this prospective paper, we tried to review the field of creativity studies from a social and cognitive standpoint and tie parallel links to the current development of AI models. Doing so, we aimed to provide a novel look at creativity in the era of artificial intelligence. After underlining the major limitations of current models from the lens of cognitive studies, we discussed how the highly contextual aspects of creativity and the question of quality measurements prove to be crippling to the \textit{singularity-seeking} approach. Hence, blindly following the aspiration of \textit{self-contained} generative machines, based on purely mimicking human creative traits appear as a somewhat pointless endeavor. Instead, we discussed the possibility of redefining the relationships between human and machines through \textit{co-creativity} approaches. This allows to scrutinize and empower the \textit{link} between both agents in the creative process, rather than the agents taken in isolation, as a main object of study.

\section{Acknowledgements}

This work is supported by the ANR MAKIMOno (17-CE38-0015-01) project, the SSHRC ACTOR (895-2018-1023) Partnership, Emergence(s) ACIDITEAM from Ville de Paris and ACIMO project from Sorbonne Universit\'e.

\bibliography{main}

\end{document}